\documentclass[sigconf]{acmart}
\usepackage{booktabs}
\usepackage{graphicx}
\usepackage[tight,footnotesize]{subfigure}
\usepackage{url}
\usepackage[utf8]{inputenc}
\usepackage{natbib}
\usepackage[font={small}]{caption}
\setcopyright{rightsretained}

\copyrightyear{2018}
\acmYear{2018}
\setcopyright{none}
\acmConference[CryBlock'18]{1st Workshop on Cryptocurrencies and Blockchains for Distributed Systems }{June 15, 2018}{Munich, Germany}
\acmBooktitle{CryBlock'18: 1st Workshop on Cryptocurrencies and Blockchains for Distributed Systems , June 15, 2018, Munich, Germany}
\acmDOI{10.1145/3211933.3211950}
\acmISBN{978-1-4503-5838-5}
    
\begin{document}

\title{A Blockchain-based Flight Data Recorder\\ for Cloud Accountability}

\author{Gabriele D'Angelo}
\orcid{0000-0002-3690-6651}
\affiliation{%
  \institution{University of Bologna}
  \city{Bologna}
  \country{Italy}
}
\email{g.dangelo@unibo.it}

\author{Stefano Ferretti}
\orcid{0000-0002-1911-4708}
\affiliation{%
  \institution{University of Bologna}
  \city{Bologna}
  \country{Italy}
}
\email{s.ferretti@unibo.it}

\author{Moreno Marzolla}
\orcid{0000-0002-2151-5287}
\affiliation{%
  \institution{University of Bologna}
  \city{Bologna}
  \country{Italy}
}
\email{moreno.marzolla@unibo.it}

\begin{abstract}
  Many companies rely on Cloud infrastructures for their computation,
  communication and data storage requirements. Whi\-le Cloud services
  provide some benefits, e.g., replacing high upfront costs for an IT
  infrastructure with a pay-as-you-go model, they also introduce
  serious concerns that are notoriously difficult to address. In
  essence, Cloud customers are storing data and running computations
  on infrastructures that they can not control directly. Therefore,
  when problems arise -- violations of Service Level Agreements, data
  corruption, data leakage, security breaches -- both customers and
  Cloud providers face the challenge of agreeing on which party is to
  be held responsible. In this paper, we review the challenges and
  requirements for enforcing accountability in Cloud infrastructures,
  and argue that smart contracts and blockchain\footnote{\textbf{WARNING}: this paper may contain traces of silicon snake oil and palm oil.} technologies might
  provide a key contribution towards accountable Clouds.
\end{abstract}



\maketitle

\section{Introduction}

According to the Cloud computing paradigm, computing resources are
viewed as a utility that is provided to customers (end users,
organizations) as required. One of the most prominent advantages of
Cloud computing is the possibility for the customers to get the
resources they need without the huge upfront and long-term investment
that would be necessary to acquire and manage the resources on their
premises. However, this means that customers do not own, and have no
direct control on, the resources they use; a common joke suggests that
the term ``Cloud computing'' should be replaced with ``other people's
computers'' so that the sentence ``storing data in the Cloud'' becomes
``storing data on other people's computers''.

Corporations and government organizations are well aware of this
issue, and the more privacy-conscious of them are willing to put only
their less sensitive data in the Cloud. This attitude limits the full
potential of the Cloud, but is clearly justified by the fact that
organizations are uncomfortable with the idea of storing their data on
systems they do not control. The legal implications of data and
applications being held by a third party, possibly in a different
country with a different data protection legislation, are complex and
not well understood. If something goes wrong (e.g., data is lost, or
the computation returns an incorrect result), how do we determine
whether the customer or the provider caused the problem? For example,
how could the following disputes be resolved?

\paragraph*{Scenario 1}
A company chooses to offload an important customer-facing application
to a Cloud infrastructure; however, the application crashes and
customers complain with the company asking for compensation. The
company blames the Cloud provider, who in turn asserts that its
infrastructure worked as expected.
  
\paragraph*{Scenario 2}
A company stores sensitive data on the Cloud. After some time, the
company discovers that the data became known to a competitor.  The
company believes that there has been a security breach on the Cloud;
the service provider denies any responsibility and refuses further
investigation.

\paragraph*{Scenario 3}
A company stores important data on the Cloud. After some time, part of
the data is missing. The company blames the Cloud provider, who
asserts that the allegedly missing data have never been there.\medskip

Cloud providers offer services on an as-is and as-available basis,
subject to terms and conditions that disclaim any responsibility no
matter what. For example, the terms and conditions for Google~Docs are
full of obligations on the user, but do not promise much in
return. Here is an
excerpt\footnote{\url{https://tools.google.com/dlpage/res/webmmf/en/eula.html},
  Accessed on 2018-01-23}:

\begin{quotation}
\textit{In particular, Google, its subsidiaries and affiliates, and its licensors do not represent or warrant to you that:\\
  (a) your use of the services will meet your requirements,\\
  (b) your use of the services will be uninterrupted, timely, secure or free from error,\\
  (c) any information obtained by you as a result of your use of the services will be accurate or reliable, and\\
  (d) that defects in the operation or functionality of any software provided to you as part of the services will be corrected.}
\end{quotation}

Amazon Web Services are no different: their general customer agreement
states that\footnote{\url{https://aws.amazon.com/agreement/}, Accessed
  on 2018-01-24}

\begin{quotation}
\textit{Further, neither we nor any of our affiliates or licensors will be responsible for any compensation, reimbursement, or damages arising in connection with: [...] (d) any unauthorized access to, alteration of, or the deletion, destruction, damage, loss or failure to store any of your content or other data.}
\end{quotation}

A little different, but still insufficient, is the agreement offered
by Google to its Cloud Storage customers, in which some service uptime
thresholds are
defined\footnote{\url{https://cloud.google.com/storage/sla/}, Accessed
  on 2018-02-27}.

In the absence of solid evidence, it would be impossible to settle
disputes. To address this concern, Cloud services need to be made
accountable~\cite{10.1007/978-3-319-61204-1_31,Haeberlen:2010,Neisse:2017:BAD:3098954.3098958,Shafagh:2017:TBA:3140649.3140656}.
Accountability is fundamental to developing trust in services. All
actions and transactions should be ultimately attributable to some
user or agent. Accountability brings greater responsibility to the
users and the authorities, while at the same time holding services
responsible for their functionality and behavior.

In this paper, we analyze the problem of enforcing accountability and
trust on Cloud infrastructures. One of the key aspects of an
accountable Cloud is a component that is responsible for logging
events in a trusted, tamper-proof way. So far, building such a
component has been highly nontrivial without resorting to a trusted
third party or to tamper-proof hardware devices. The blockchain
technology might change this, allowing the implementation of
distributed, unforgeable event logs. Additionally, the blockchain
allows the implementation of \emph{smart contracts}~\cite{Szabo:1997}
through which it might be possible to write programs that can
negotiate and verify the fulfillment of Service Level Agreements
(SLAs).

This paper is organized as follows. In Section~\ref{sec:background} we
provide some background on Cloud computing, blockchain and smart
contracts. In Section~\ref{sec:cfr} we highlight some of the
challenges and requirements of accountable
Clouds. Section~\ref{sec:case} investigates how blockchain-based
technologies can help to address the challenges above in a case study
dealing with accountable cloud-based storage. Finally, conclusions and
future research directions are discussed in
Section~\ref{sec:conclusions}.

\section{Background}\label{sec:background}

To make this paper self-contained, we provide some background on Cloud
computing infrastructures, accountability, blockchain technology and
smart contracts.

\subsection{Cloud Computing}

The essential characteristics of a Cloud environment can be summarized
as follows~\cite{NIST}: \emph{on-demand self service} refers to the
ability to provide resources (e.g., CPU time, network storage) as
needed~\cite{Armbrust:2010,Marzolla:2012}; \emph{broad network access}
means that resources can be accessed through the
network~\cite{Armbrust:2010}; \emph{resource pooling} requires that
virtual and physical resources can be pooled and assigned dynamically
to consumers using a multi-tenant model~\cite{Marzolla:2012};
\emph{elasticity} is the ability of dynamically provisioning resources
to enable customer applications to scale up and
down~\cite{Armbrust:2010,Marzolla:2012}; fimally, \emph{measured
  service} means that Cloud resource and service usages are optimized
through a pay-per-use model~\cite{Ferretti:2010,Haeberlen:2010}.

\subsection{Accountability in Cloud Computing}

The importance of accountability in distributed systems in
general~\cite{Haeberlen:2007,Yumerefendi:2004} and Cloud computing in
particular~\cite{Santos:2009, Haeberlen:2010} has already been
recognized. In~\cite{Haeberlen:2010} the author discusses the
requirements for achieving accountability in clouds through
tamper-evident logs: \emph{completeness} (all~SLA violations are
eventually reported); \emph{accuracy} (no violations are reported if
the~SLA is not violated); \emph{verifiability} (all reported
violations can be independently verified by a third party).

To actually realize an accountable Cloud based on trusted logs it is
necessary to decide \emph{what} to log and \emph{how} to log.  We
consider 'how' first. Logging must guarantee fairness and
non-repudiation, ensuring that well-behaved parties are not
disadvantaged by the misbehavior of others, and that no party can
subsequently deny their participation. It should enable tracing back
the causes of an `incident' (i.e., a behavior that is not SLA
compliant) after it has occurred.  Cloud providers and customers
require protection with respect to each other's actions, with provider
assurances rooted in an independent source of trust. For example,
there should be user-verifiable assurance that the data, applications
and services they deploy in the Cloud are secure even against
compromise by Cloud system administrators.

As concerns `what' to record, Cloud computing creates new
relationships between an organization and third party Cloud service
providers. The general scenario is that Cloud services could be
arbitrarily complex. Providers will offer their services to consumers
with specific Quality of Service~(QoS) attributes, such as
reliability, security, under specific terms and
conditions~\cite{Ferretti:2010}. Most of the existing research on~SLA
management focuses on computational and algorithmic aspects of~QoS
monitoring and provisioning. Specifically, considerable effort has
been spent in developing proactive or reactive algorithms for
allocating the appropriate number and kind of resources needed to meet
a set of~QoS requirements.  However, SLA violations do happen in
practice, and it is necessary to deal with them. Currently, SLA
violations must be handled entirely on the basis of ``out of band''
negotiations between service providers and customers, since the
systems being monitored are unable to provide legal evidences of
malfunctions (or lack of).  The lack of a well-defined framework for
identifying violations and assigning responsibilities is a limiting
factor.

\subsection{Blockchain}

A blockchain is a distributed ledger that records transactions in
blocks~\cite{Antonopoulos:2014:MBU:2695500}. Each block contains a set
of transactions and it has a link to a previous block, thus creating a
chain of chronologically ordered blocks. Transactions within a block
are assumed to have happened at the same time. In the typical
scenarios, transactions record an exchange of digital currencies, but
in fact they can be employed to record any kind of event.

What makes the blockchain technology appealing is that the combination
of peer-to-peer systems, cryptographic techniques, use of distributed
consensus schemes and pseudonimity ensure that the set of confirmed
transactions becomes public, traceable and tamper-resistant.  The
latter property is obtained by linking subsequent blocks together
using cryptographic hash functions so that the modification of
transaction data on a block~$B_i$ would change the hash that is
contained in the subsequent block~$B_{i+1}$, thus altering the content
of block $B_{i+1}$ and so on. The blockchain is replicated across
multiple nodes in a peer-to-peer fashion: therefore, any attempt to
alter the blockchain would create an easily detectable inconsistency
of all replicas.

The blockchain uses digital pseudonyms -- usually, a hash of an
address -- to provide some level of anonymity. Therefore, everyone can
trace the activities of an entity with a given pseudonym, but it is
computationally expensive (although not impossible) to associate a
pseudonym back to a specific entity or individual. This property
further contributes to make the blockchain an interesting tool to
build a tamper-proof log to be used in accountable Clouds.

\subsection{Smart Contracts}

The concept of \emph{Smart Contract} was developed by
Szabo~\cite{Szabo:1997}. A smart contract is a program representing an
agreement that is automatically executable and enforceable by nodes
that participate in the blockchain management. The automatic execution
of the program is triggered when certain conditions are met, and the
program deterministically executes the terms of a contract, specified
as software code. Examples of implementations are from
Ethereum~\cite{Ethereum} and IBM Hyperledger~\cite{Cachin:2016}.

An interesting aspect of smart contracts is their ability to be self-enforcing in the verification of the
fulfillment of~SLA agreements in a Cloud computing environment.

Smart contracts provide ways
of formulating machine-readable sets of rules from service contracts, thus transforming in software code some rules that are typically written in
``legal-ese''.
In our scenario, smart contracts might be set to contain two kinds of
contractual clauses: (\emph{i}) terms and conditions and (\emph{ii}) SLAs.  Terms and
conditions are concerned with rights, obligations and prohibitions to
perform a particular action;
whereas~SLAs are concerned with right, obligations and
prohibitions to maintain a given service in a particular state.

Smart contracts can represent the basis of systematically determining
monitoring requirements for detecting rule violations. This can be
accomplished by recording service interactions at a granularity that
is sufficient for checking if they comply with the rights
(permissions), obligations and prohibitions stipulated in contract
clauses and tracing causes of violations.

\section{A Blockchain based Proposal}\label{sec:cfr}

In this section we discuss on the viability of employing blockchain
technologies as a ``Cloud flight data recorder'', in order to log all
interactions among different parties and record them in a trusted,
tamper-proof way. These interactions, represented as transactions
recorded in the blockchain, can be checked by all the involved parties
or by smart contracts, and can be used to solve disputes arising due
to~SLA violations.

The basic idea is that all operations accomplished in the Cloud are
recorded in the blockchain. Instantiation of a virtual machine,
upload, deletion or modification of files, access timestamp of a given
resource, are all examples of events that can be recorded. All these
events must be notified in the blockchain either by the entity
invoking the request, e.g., the user (or his delegate) that asks
access to a file, and/or by the entity receiving the request, i.e.,
the Cloud provider. The rationale is that recording all the activities
of the involved parties can help to reveal the causes of a~SLA
violation.

As an example, let us consider a Cloud storage service for data
archival and backup, such as Amazon Glacier. This service allows users
to (\emph{i}) store a data block~$x$, (\emph{ii}) delete~$x$,
(\emph{iii}) read back~$x$. In this case, the blockchain can be used
as a flight data recorded, following a notary scheme. Let us assume
that the provider cannot deliver a data block~$x$ requested by the
user, or the the provided data is different than what expected. In
this case, inspection on the blockchain can reveal whether the
provider lost~$x$ or some updates to~$x$, or the user has deleted or
never uploaded~$x$ (or some modifications to~$x$).

Another typical example of SLA in Cloud computing services is: ``99\%
of transactions during a daily activity must have a response time
below a certain time~$t$''.  In this case, we can assume the presence
of a (third) trusted software/hardware component that logs response
times or can audit (virtualized) resource
usage~\cite{Jayathilaka:2017}. In blockchain terminology, tamper-proof
trusted entities that use a secure channel to send data to a smart
contract or to the blockchain are referred to as
``oracles''~\cite{bashir2017mastering}. As oracles populate the
blockchain, we can envision self-enforcing smart contract that
monitors response times and, based on the SLA, pays the provider
accordingly.

A similar strategy could be exploited if one decides to monitor SLAs
stipulated in terms of effective resource capacities provided by the
Cloud, rather than applications-specific performance
metrics~\cite{Kesidis:2016}. Thus, the Cloud flight data recorder
would allow checking if the Cloud provider allocated processing and
storage capacities, RAM, middleware resources.

\section{Case study}\label{sec:case}

In this section we describe a possible implementation of a simple
accountable Cloud-based storage in which the execution of some
specific operations leads to the verification of a given~SLA
agreement. For example, a customer uploads some content on a Cloud
storage service. Let us assume for simplicity that the content is a
file (a similar reasoning would apply to data chunks or other kinds of
information). In the following, the customer wants to be sure that the
uploaded files are not removed or altered by the Cloud provider. This
can be obtained by using different execution architectures:
\begin{enumerate}
  \item blockchain-based double signed transactions;
  \item blockchain-based logging (without smart contracts);
  \item blockchain-based logging and smart contracts.
\end{enumerate}

Double signed transactions have the peculiarity of being signed by
multiple parties. Thus, they can certify a transaction that has been
agreed by both the customer (User) and the cloud provider
(Cloud). Double signed transactions are a straightforward tool to be
used in a blockchain and require a low overhead since it can be
realized with few interactions. On the other hand, this approach would
provide a coarse-grained representation of the interactions between
the User and the Cloud. In fact, such an approach certifies whether
the parties agreed on something. This all or nothing result can be
quite limiting since it relies on the two parties to actually agree.

Using a blockchain-based logging (without smart contracts) permits
recording all the interactions between the User and the Cloud. In case
of~SLA violations, each party can trigger a verification by a third
entity (e.g., an arbitrator) to identify who is responsible for such a
violation. It is worth noting that, in this architecture, the
arbitrator is not required to have been involved in any previous
interaction with either the User or the Cloud, since it can use the
information publicly provided by the blockchain to determine the
responsibilities.

Another option is to employ a smart contract acting as the
arbitrator. In this approach, the smart contract is in charge of
verifying all events stored in the blockchain, identifying~SLA
violations and calculating compensations. The main advantage of this
approach is that no third party needs to be involved to resolve
disputes. In particular, since the content of the smart contract can
be accessed by both parties, they can verify its correctness before
agreeing on its terms. In other words, the trust of the User and the
Cloud provider is on the smart contract (that can be inspected and
verified), following the notion that ``code is law''.

In the following, we describe a simple protocol that can help identify
responsibilities on~SLA violations for the following three operations:
upload, delete, read to a file. We assume that the following active
entities are involved: User, Cloud provider, the arbitrator (here
referred as Smart-contract). To simplify the discussion, we also
consider the Blockchain as a passive entity, capable of receiving and
storing events generated by active entities. For the sake of a simpler
description we might state that an entity, let say the Cloud, receives
a transaction from the Blockchain. This is actually a simplification
to state that the nodes working on the blockchain reached a consensus
on a transaction that involves the~Cloud entity as the receiver. Thus,
the transaction has been inserted into a valid block.

We also assume that each data file is encrypted by the User before
uploading it on the Cloud. The Cloud is not able to decrypt the
file. Thus, in case of an insider threat, the malicious entity would
only read a ciphertext.

\begin{figure}[h]
\centering
\includegraphics[width=8.5cm]{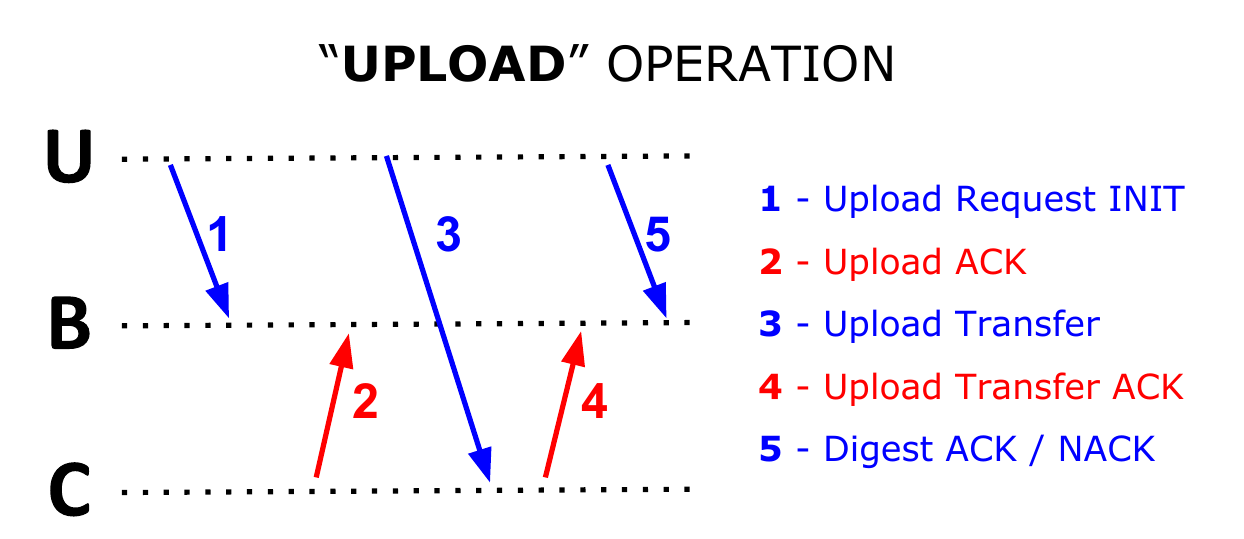}
\caption{Upload of a data from a User (U) to the Cloud (C). Arrows
  between involved entities and the Blockchain (B) represent
  transactions inserted in the blockchain, e.g., the arrow from~U to~C
  represents data transmission from the User to the Cloud.}
\label{fig_upload}
\end{figure}

Figure~\ref{fig_upload} shows the behavior of the involved entities
when the User uploads a file. Before transmitting the data to the
Cloud, it registers in the Blockchain the upload request
initialization (arrow~$1$ in the figure). This request is in fact a
transaction, stored in the blockchain, from the User to the Cloud;
recall that the content of the blockchain is public, so the Cloud can
see what the User stored in the blockchain. Once the Cloud receives
the transaction from the (nodes that agreed on the transaction
inserted into the) blockchain, it can accept the upload request by
issuing an upload ACK transaction to the User (arrow~$2$). Once the
User receives the ACK transaction, it can start the data upload to the
Cloud (arrow $3$). Once the upload finishes, the Cloud logs the
success of this operation with a new transaction (arrow $4$); in this
transaction, the Cloud stores in the blockchain the digest of the data
it received. Then, the User acknowledges the end of the upload (arrow
$5$); in turn, the User confirms (rejects) the digest published by the
Cloud with a digest ACK (NACK). In this way, anyone (i.e.,~an
arbitrator) can verify the correctness of the uploaded data, by
checking the digest provided by the Cloud and the related confirmation
by the User. If the User rejects the Cloud's digest, the Cloud should
delete the received data.  In these operations, the arbitrator is not
involved in the process. However, in case of a dispute, it can check
all the transactions and understand if one of the two parties did not
behave correctly.

\begin{figure}[h]
\centering
\includegraphics[width=8.5cm]{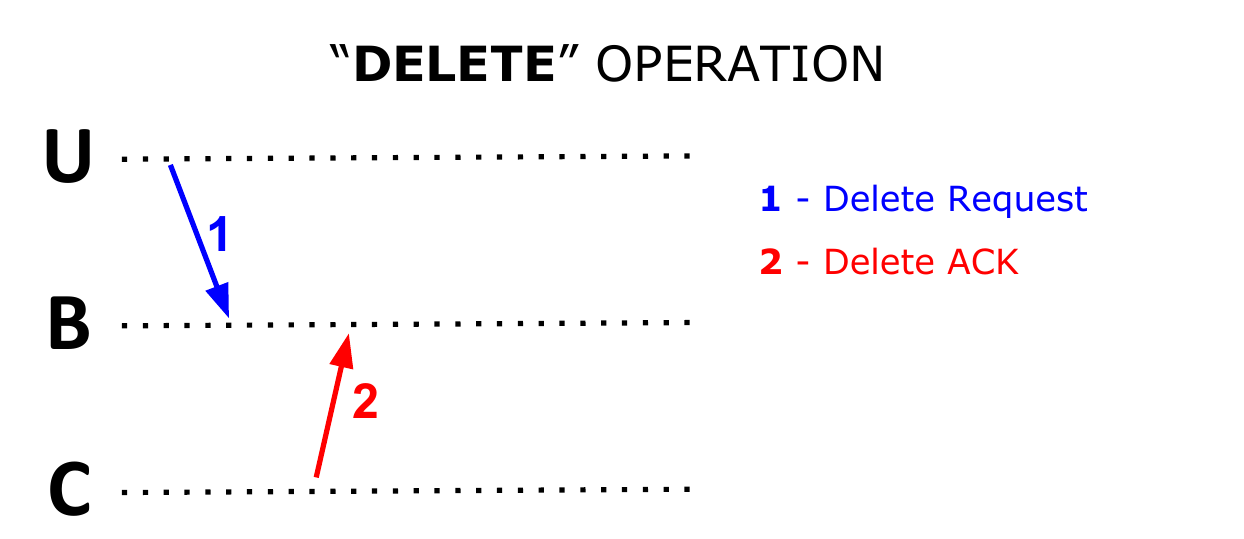}
\caption{Deletion of a data from a User (U) to the Cloud (C). Arrows from involved entities to the Blockchain (B) represent transactions inserted in the blockchain.}
\label{fig_delete}
\end{figure}

The interactions required to delete a file from the Cloud are shown in
Figure \ref{fig_delete}. The User issues a delete request by creating
a related transaction to the Cloud and inserting into the Blockchain
(arrow $1$). As a consequence, the Cloud will receive this
transaction, deletes the data and acknowledges this deletion by
registering the event into the Blockchain (arrow $2$).  After that,
future disputes on the presence (absence) of a data can be regulated
by looking at the log. In fact, if the User requests data that is
not present in the Cloud, it is possible to verify whether the User
previously asked a deletion for that data. If a related transaction is
present, the Cloud correctly deleted that file; if not, we are in
presence of a~SLA violation. Additionally, if during a dispute the
Cloud is found to have a copy of a file that the User asked to delete,
and acknowledged delete operation is in the blockchain, then the Cloud
might be held responsible of a~SLA violation since it did not properly
removed the file as requested.

\begin{figure}[h]
\centering
\includegraphics[width=8.5cm]{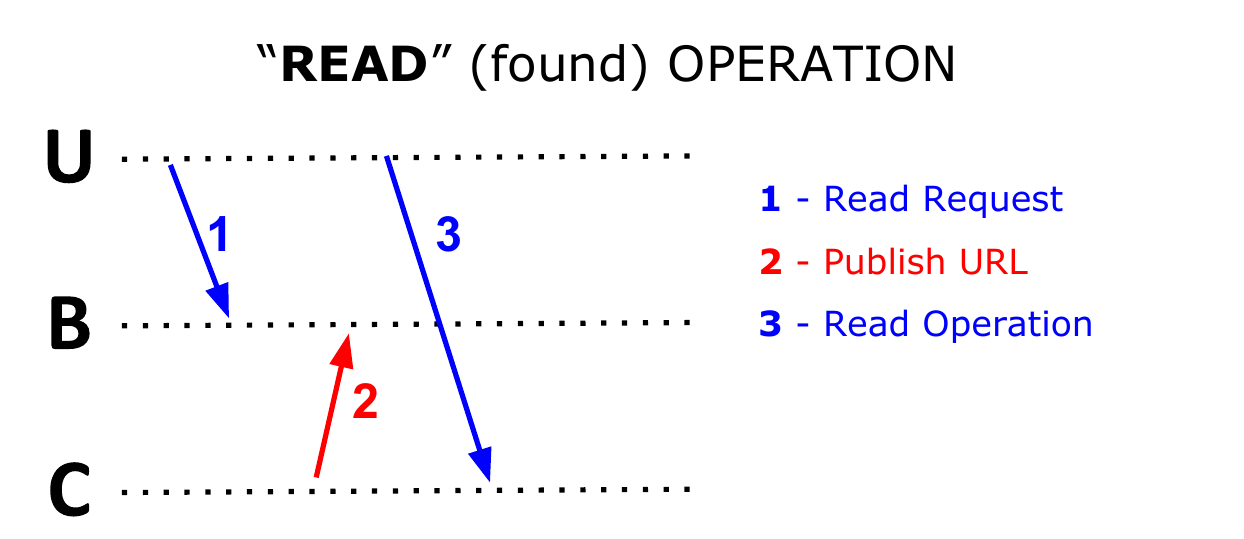}
\caption{Successful read of a data requested by the User (U) to the Cloud (C). Arrows from involved entities to the Blockchain (B) represent transactions inserted in the blockchain; the arrow from U to C represents the access to the data.}
\label{fig_read_found}
\end{figure}

Figure~\ref{fig_read_found} shows the interactions required to read a
file stored in the cloud.  In this case, the tricky part is to track
access of a data by all users. We notice that, since we assume that
the data stored into the Cloud is encrypted, only the authorized
parties can decrypt it and gain access to the actual content. This
prevents the Cloud to send sensible information to non-allowed
parties. In order to access a file, the User issues a transaction to
the Cloud representing a read request (arrow $1$). To give access to
the data, the Cloud inserts into the blockchain a URL, where the file
can be retrieved. This procedure is required in order to witness the
fact that the Cloud has granted access to the User and that the file
is the valid one. Indeed, anyone (i.e., the arbitrator) can verify the
content of the URL, without accessing the real data (that is
encrypted).  Thus, this procedure allows also comparing the digest of
the provided data with that stored in the blockchain, in order to
understand if the provided data to read complies with that previously
uploaded.

\begin{figure}[h]
\centering
\includegraphics[width=8.5cm]{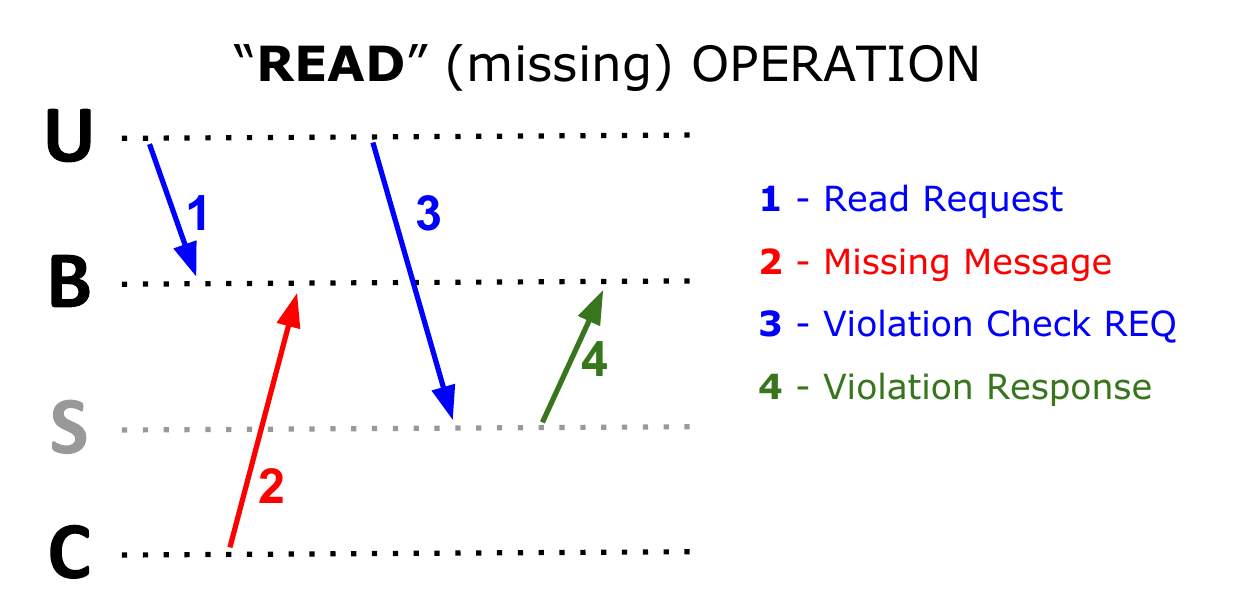}
\caption{Unsuccessful read of a data requested by the User (U) to the Cloud (C). Arrows $1$ and $2$ from involved entities to the Blockchain (B) represent transactions inserted in the blockchain; the arrow $3$ from U to S represents the triggering of the Smart-contract (S). Arrow $4$ represents the output of (S) that is stored on the (B).}
\label{fig_read_missing}
\end{figure}

Figure~\ref{fig_read_missing} shows the interactions among the
entities when the data requested by the User is missing from the
Cloud. As before, the user issues a transaction to the Cloud
representing a read request (arrow $1$). The Cloud verifies that the
request data is missing from its storage and responds with a missing
message (arrow $2$). To assess if there is a SLA violation and its
attribution, the User triggers the Smart contract (arrow $3$). By
analyzing the transaction history on the blockchain, the Smart
contract can determine if a SLA violation has occurred and, for
example, determines the related compensation. For obvious reasons, the
output of this process is stored on the blockchain (arrow $4$). In
alternative, the User intervention can be avoided implementing a Smart
contract that monitors the events on the blockchain and that
self-activates when necessary. Clearly, this approach would increase
the cost of running the Smart Contract.

The case in which a data stored on the Cloud is not missing but it has
been altered (e.g. failed digest check) is very similar to the failed
read. More generally, the proposed architecture can support other
(more complex) SLAs on the services provided by the cloud provider
that are not discussed in this paper.

\section{Conclusions}\label{sec:conclusions}

In this paper we explored the use of blockchain technologies to build
a ``flight data recorder'' for Cloud accountability. The blockchain
allows pseudo-anonymous and tamper-proof logging of events into a
distributed ledger. The ledger can then be used to verify if~SLAs are
violated. Moreover, the presence of self-enforcing smart contracts
allow to automatically identify responsibilities and settle disputes,
for instance making automatic payments based on the offered service.

An issue that needs further investigation is that of
efficiency. Indeed, the current incarnations of the blockchain might
not provide a response time short enough to efficiently support all
the interactions shown in Section~\ref{sec:case} from a large number
of customers operating concurrently. Additionally, transaction fees
might represent an economic disincentive to the above-mentioned
approach. Thus, the choice of which blockchain technology to use
remains an important problem. Probably, a traditional Bitcoin-like
blockchain solution would not be the most appropriate in this
context. Instead, a permissioned blockchain would have the advantage
of being more performant, scalable, and only accessible by a dedicated
group of entities, which has the eligibility to join it. Lightweight,
permissionless blockchains with low or no-fees transactions exists,
e.g., IOTA\footnote{\url{https://iota.org/}, accessed on
  2018-03-02}, but unfortunately do not yet support smart contracts.

\bibliographystyle{ACM-Reference-Format}
\bibliography{cloud-accountability.bib}

\end{document}